# Microstructure reconstruction using diffusion-based generative models

Kang-Hyun. Lee[1], and Gun Jin Yun[1,2]

[1]*Department of Mechanical & Aerospace Engineering, Seoul National University, Gwanak-gu Gwanak-ro 1 Seoul 08826, Republic of Korea*
[2]*Institute of Advanced Aerospace Technology, Seoul National University, Gwanak-gu Gwanak-ro 1, Seoul 08826, South Korea*

**Abstract**

Microstructure reconstruction has been an essential part of computational material engineering to reveal the relationship between microstructures and material properties. However, finding a general solution for microstructure characterization and reconstruction (MCR) tasks is still challenging, although there have been many attempts such as the descriptor-based MCR methods. To address this generality problem, the denoising diffusion models are first employed for the microstructure reconstruction task in this study. The applicability of the diffusion-based models is validated with several types of microstructures (e.g., polycrystalline alloy, carbonate, ceramics, copolymer, fiber composite, etc.) that have different morphological characteristics. The quality of the generated images is assessed with the quantitative evaluation metrics (Fréchet inception distance (FID) score, precision, and recall) and the conventional statistical microstructure descriptors. Furthermore, the formulation of implicit probabilistic models (which yields non-Markovian diffusion processes) is adopted to accelerate the sampling process, thereby controlling the computational cost considering the practicability and reliability. The results show that the denoising diffusion models are well applicable to the reconstruction of various types of microstructures with different spatial distributions and morphological features. The diffusion-based approach provides a stable training process with simple implementation for generating visually similar and statistically equivalent microstructures. In these regards, the diffusion model has great potential to be used as a universal microstructure reconstruction method for handling complex microstructures for materials science.

## 1. Introduction

One of the major goals in material science is to build process-microstructure-property linkages of the target material to optimize its property for the given application. Owing to the development of computational materials science with integrated computational materials engineering (ICME), inverse materials design has been aided by numerical simulations and data-driven techniques in the last decades [1-6]. A vital issue in the computational material design framework is building a reliable large-scale microstructure database. To deal with this issue, microstructure characterization and reconstruction (MCR) have been utilized to facilitate the materials design by using microstructural descriptors which encapsulate the essential morphological information. For instance, if one can find a reasonable descriptor for a given microstructure image, it is able to reconstruct equivalent microstructure images (or even three-dimensional microstructure) as many as desired.

To date, a considerable amount of studies have been conducted to propose some reasonable and universal descriptors for microstructure reconstruction such as physical descriptors [3, 7], correlation functions [8, 9], random fields [10-12], and spectral density functions [13, 14]. Based on the pre-characterized microstructure with certain descriptors, most of the MCR methods involve an iterative optimization process for minimizing the difference between the target descriptors and the current descriptors [1, 15-18]. After a sufficient number of iterations, it is able to generate statistically and morphologically equivalent microstructure samples. In addition, the differentiable descriptors have been also used to characterize and reconstruct microstructure samples with gradient-based optimization [3, 4, 7, 19]. However, it is still challenging to find a universal descriptor that works for various types of microstructures. The conventional MCR methods based on statistics of morphological features (e.g., two-point correlation and lineal path functions) often do not work well as the shape of the target microstructure gets complicated. For instance, the typical microstructures of alloy, ceramics, and carbonate are well reconstructed from n-point correlation [3, 8] whereas only some artificial neural network (ANN)-based models [3, 4, 20] have shown noteworthy performance on the reconstruction of copolymer microstructure. In other words, the conventional MCR methodologies require insights into the spatial distribution of the morphological features in the material of interest.

Meanwhile, various types of data-based synthesis frameworks have been proposed in recent days with the rapid evolution of generative neural network models including variational autoencoders (VAEs) [21, 22], flow-based models [23-25], and generative adversarial networks (GANs) [5, 26-28]. These generative models learn a latent representation of the sample data, which can also be thought of as a compressed form of original data. After training the generative models, the models can generate a new sample by simply sampling the latent variables. In particular, GAN-based models have shown noteworthy performance on microstructure reconstruction tasks [5, 20, 23, 26, 29, 30]. In addition, the StyleGAN [31, 32] also has the lowest Fréchet inception distance (FID) scores with ImageNet $64 \times 64$ dataset as shown in Table 1. Using GAN, it is able to generate high-resolution image samples with high precision and recall (i.e., low FID). Furthermore, not only do GANs generate high-quality samples, GANs can generate samples very fast since it requires only a single pass through the trained network. However, GAN models suffer from critical problems such as mode collapse and non-convergence issues [27, 33-35]. Due to the adversarial formulation of the loss function to train the discriminator and the generator, training GANs is highly unstable [36, 37]. For instance, GAN models often produce samples from a limited set of modes (i.e., degraded diversity of the model outputs). To secure the diversity of samples as well as the precision, the hyperparameters and the complex features in GAN models should be carefully addressed. The difficulties of training and scaling models are the major drawback of GANs, which hinder their wide application in different domains.

To overcome these problems with GANs, diffusion-based generative models have been drawing significant attention for data synthesis in recent days. Recent works on diffusion models have demonstrated their noteworthy generative performance with high-quality samples comparable to those from GANs [38-41]. In 2021, Dhariwal and Nichol [38] also showed that their tailored denoising diffusion models outperform GANs on image synthesis tasks in sample quality (i.e., FID score). In addition, the performance of the diffusion-based model (i.e., cascaded diffusion model (CDM)) presented by Ho et al. [42] is also ranked first with the benchmark result using ImageNet $64 \times 64$ (Table 1). The diffusion models are likelihood-based models which generate samples by gradually denoising the noised samples [38-41]. According to Yang et al. [43], there are three popular formulations of diffusion models which are the denoising diffusion probabilistic models (DDPM) [41], the score-based generative models (SGM), and the stochastic differential equations (SDE). Although the detailed formulation process is different for each case, DDPM and SGM can be generalized with the solutions of SDE for noise perturbation and sample generation (i.e., denoising). In particular, the open-source code for the generative model formulated with the DDPM approach is also released by OpenAI [39, 41]. After the original denoising diffusion model was proposed by Sohl-Dickstein [44], the diffusion models have been shown to provide a generative performance with a comprehensive distribution coverage in various domains such as computer vision, natural language processing, and multimodal modeling [43, 45]. Furthermore, the stationary objective of these models without adversarial formulation and relatively simple architecture compared to GANs also facilitate their application. The advances in the probabilistic parameterization of the diffusion models also provide stable training procedures for forward (diffusion) and backward (denoising) transformation of the prior data distribution [39, 46].

This study proposes a diffusion model-based microstructure reconstruction approach using the DDPM formulation [39, 41]. Compared to the conventional MCR methods, the model does not require any physical or spatial approximation. Instead, the model directly learns the sample data distribution, enabling the generation of multiple synthetic images of microstructures. To the authors' best knowledge, this is the first study conducted for implementing denoising diffusion-based microstructure reconstruction. Furthermore, the implicit formulation of probabilistic diffusion models, which is called denoising diffusion implicit models (DDIM) [40, 47], is adopted to accelerate the sampling process (which is a solution for the slow computational speed of the original DDPM). To validate the model's performance, various microstructure images with different morphological characteristics are considered to be reconstructed. Then, the performance of the model is evaluated using the evaluation metrics (FID scores [48], precision, and recall [49]) for assessing the quality of samples in terms of data synthesis, as well as the conventional spatial statistics for analyzing the morphological similarities between the original and generated samples.

Table 1. FID score benchmark for image generation on ImageNet $64 \times 64$

| Model | FID | Type | Year | References |
|---|---|---|---|---|
| CDM | 1.48 | Diffusion | 2021 | [42] |
| StyleGAN-XL | 1.51 | GAN | 2022 | [31] |
| ADM | 2.07 | Diffusion | 2021 | [38] |
| Improved DDPM | 2.92 | Diffusion | 2021 | [39] |
| PGMGAN | 21,73 | GAN | 2021 | [50] |
| GLIDE+CLIP+CLS+CLS+FREE | 29.184 | Diffusion | 2022 | [51] |
| GLIDE+CLS-FREE | 29.219 | Diffusion | 2022 | [51] |

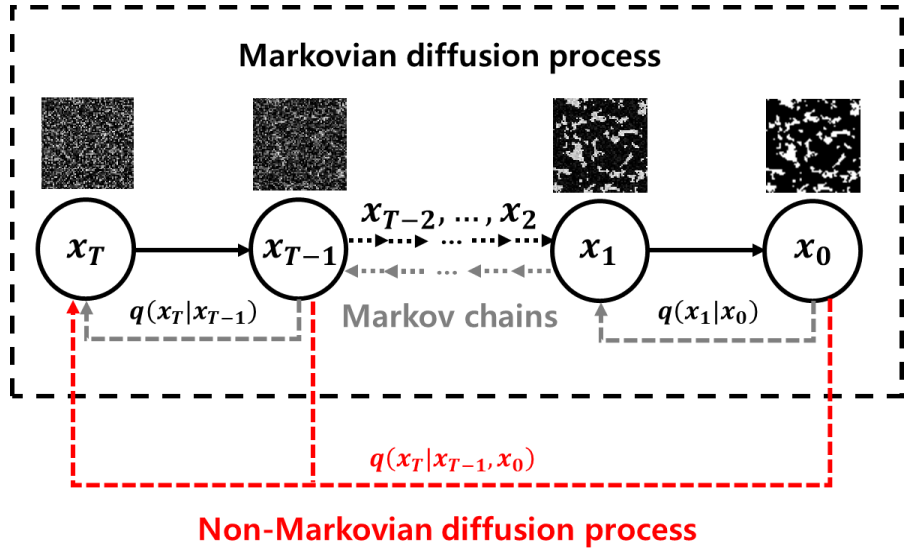

Figure 1. Schematic diagram of Markovian diffusion process and non-Markovian inference models for microstructure synthesis

## 2. Diffusion probabilistic modeling for microstructure reconstruction

2.1. Fundamentals of diffusion models

There are several different formulations for diffusion models presented in previous literature [43]. The denoising diffusion probabilistic model proposed by Sohl-Dickstein et al. [44] consists of a parameterized Markov chain that destroys structure in a data distribution by iteratively adding noise and a reverse process that restores the structure in data. Their study also demonstrated the denoising diffusion-based model's effectiveness in sampling various types of data, including complex natural image datasets. Similarly, Song and Ermon [46] proposed a diffusion-based generative model using score matching (i.e., noise-conditioned score networks) that estimates the gradients of data densities. Their result showed that the trained model could generate high-quality samples comparable to GANs based on Langevin dynamics without additional sampling process. Furthermore, Ho et al. [41] demonstrated DDPM, which is a certain parameterized version of the diffusion probabilistic model, yielding an equivalence with denoising score matching (during training) and annealed Langevin dynamics (during sampling). In this study, the DDPM [41] approach (Figure 1) is adopted to build a generative model, as well as the DDIM [40, 47] approach for accelerating the sampling process.

### 2.1.1 Formulation of DDPM

According to the diffusion models with Gaussian noise formulated by Ho et al. [41], a Markovian noising process $q$ that progressively adds noise to the sample data at different time step $t$ can be defined as

$$q(x_t \mid x_{t-1}) := N\left(x_t; \sqrt{1-\beta_t}x_{t-1}, \beta_t \mathbf{I}\right) \quad (1)$$

where $x_1, \dots, x_T$ are the noised samples (which also act as the latent variables in DDPM) from the sample data distribution $x_0 \sim q(x_0)$, and $\beta_t$ is the parameter for controlling the noising schedule. Herein, the noising process $q(x_t|x_0)$ can be expressed as a Gaussian distribution which eliminates the need for sampling from repeated noising processes as

$$q(x_t|x_0) = N\left(x_t; \sqrt{\bar{\alpha}}x_0, (1-\bar{\alpha}_t)\mathbf{I}\right) \quad (2)$$

so $x_t$ can be written as a linear combination of sample data $x_0$ and a noise variable $\epsilon \sim N(0, \mathbf{I})$ by reparametrizing as follows:

$$x_t = \sqrt{\bar{\alpha}_t}x_0 + \epsilon\sqrt{1-\bar{\alpha}_t} \quad (3)$$

$$\alpha_t := 1 - \beta_t \tag{4}$$

$$\bar{\alpha}_t := \prod_{i=0}^{t} \alpha_i \tag{5}$$

It is worth noting that as $\bar{\alpha}_t$ get close to 0, $q(x_t|x_0)$ approaches to a standard Gaussian distribution $N(0, \mathbf{I})$. Sohl-Dickstein et al. [44] also noted that $q(x_{t-1}|x_t)$ converges to a diagonal Gaussian distribution as $T \to \infty$. Since all the conditional distributions can be modeled as Gaussians, it is able to train a neural network model $p_\theta$ to predict a mean and covariance matrix for denoising purpose as follows:

$$p_\theta(x_{t-1}|x_t) := N\big(x_{t-1}; \mu_\theta(x_t, t), \Sigma_\theta(x_t, t)\big) \tag{6}$$

To learn the actual data distribution using the above model (i.e., $p_\theta(x_{t-1}|x_t) \approx q(x_{t-1}|x_t)$), the variational lower-bound (VLB) can be defined to be optimized as follows [39]:

$$L_{\text{vlb}} := L_0 + L_1 + \cdots + L_{T-1} + L_T \tag{7}$$

$$L_0 := -\log p_\theta(x_0|x_1) \tag{8}$$

$$L_{t-1} := D_{KL}\big(q(x_{t-1}|x_t, x_0) \| p_\theta(x_{t-1}|x_t)\big) \tag{9}$$

$$L_T := D_{KL}\big(q(x_T|x_0) \| p(x_T)\big) \tag{10}$$

where $D_{KL}$ is Kullback-Leibler divergence. On the other hand, DDPM can be trained to learn the true data distribution with the simplified objective defined by Ho et al. [41] using the trainable mean functions and constant variances as

$$L_{\text{simple}} := \mathbb{E}_{x_0, \varepsilon, t}\left[\left\|\epsilon - \epsilon_\theta^{(t)}\big(\sqrt{\bar{\alpha}_t}x_0 + \sqrt{1-\bar{\alpha}_t}\epsilon\big)\right\|_2^2\right] \tag{11}$$

where $\epsilon_\theta^{(t)}$ is a set of noise parameters indexed by $t$ which are predicted with trainable parameters $\theta^{(t)}$. In other words, the training objective can be defined as a mean-squared error between the predicted noise and the true noise without considering the reverse process variance $\Sigma_\theta(x_t, t)$ into VLB. It is also worth noting that Eq. (11) has no regularization term since the fixed noise scheduling can lead to isotropic Gaussians with a sufficient number of diffusion time steps [38, 41, 43, 45]. Furthermore, the loss function $L_{\text{simple}}$ is equivalent to the variational inference objective in DDIM if a certain weight factor is applied [40]. In other words, the objective in DDPM can be employed as a surrogate objective for DDIM.

Meanwhile, it can be seen that Eq. (11) does not incorporate terms related to log-likelihood. However, Ho et al. [39] showed that the model could be improved by learning $\Sigma_\theta(x_t, t)$. In order to consider the effect of $\Sigma_\theta(x_t, t)$ in loss function, the following hybrid objective [38, 39] is adopted

$$L_{\text{hybrid}} = L_{\text{simple}} + \lambda L_{\text{vlb}} \qquad (12)$$

where $\lambda$ is a weight factor that is set to 0.001 to treat $L_{\text{simple}}$ as a main component of loss function while suppressing the effect of $L_{\text{vlb}}$.

### 2.1.2 Sampling with DDIM

Although DDPM had demonstrated as a promising generative model for data synthesis, the major drawback of DDPM is the high computational cost due to the many iterations. If a single sample needs to be generated with a trained DDPM, the reverse of the forward diffusion process (i.e., noising process) needs to proceed with thousands of iterative steps. This is the major difference compared to GAN where only a single pass through a trained network is required to generate a sample.

To accelerate the sampling process with DDPM, a sampling approach based on the implicit probabilistic models which is called DDIM [47], was proposed by Song et al. [40]. The basic idea of DDIM is to generalize the Markovian (stochastic) forward diffusion process to non-Markovian (deterministic). Unlike the definition of the process $q$ (Eq. (1)-(2)), an inference distribution $q_\sigma$ can be defined as follows:

$$q_\sigma(x_1, \ldots, x_T | x_0) = q_\sigma(x_T | x_0) \prod_{t=2}^{T} q_\sigma(x_{t-1} | x_t, x_0) \tag{13}$$

$$q_\sigma(x_{t-1} | x_t, x_0) = N(x_{t-1}; \tilde{\mu}_{t-1}(x_t, x_0), \sigma_t^2 I) \tag{14}$$

$$\tilde{\mu}_{t-1}(x_t, x_0) = \sqrt{\bar{\alpha}_{t-1}} x_0 + \sqrt{1 - \bar{\alpha}_{t-1} - \sigma_t^2} \cdot \frac{x_t - \sqrt{\bar{\alpha}_t} x_0}{\sqrt{1 - \bar{\alpha}_t}} \tag{15}$$

Then, the conditional forward diffusion process (which is also Gaussian) can be written as follows using Bayes' rule.

$$q_\sigma(x_t | x_{t-1}, x_0) := \frac{q_\sigma(x_{t-1} | x_t, x_0) q_\sigma(x_t | x_0)}{q_\sigma(x_{t-1} | x_0)} \tag{16}$$

As can be seen in Eq. (16), it can be said that the process is not Markovian since it depends on $x_{t-1}$ and $x_0$. In particular, the parameter $\sigma_t$ determines the stochasticity of the given forward process. For instance, the process becomes deterministic when $\sigma_t = 0$ and the corresponding inference model is called DDIM. On the other hand, the process becomes the DDPM if $\sigma_t = \sqrt{(1 - \alpha_{t-1})/(1 - \alpha_T)} \sqrt{1 - \alpha_t/\alpha_{t-1}}$. A sample $x_{t-1}$ from a sample $x_t$ can be generated with the following equation [40].

$$x_{t-1} = \sqrt{\alpha_{t-1}} \left( \frac{x_t - \sqrt{1 - \alpha_t} \epsilon_\theta^{(t)} x_t}{\sqrt{\alpha_t}} \right) + \sqrt{1 - \alpha_{t-1} - \sigma_t^2} \cdot \epsilon_\theta^{(t)} x_t + \sigma_t \epsilon_t \tag{17}$$

In particular, the major advantage of DDIM is that it is unnecessary to reverse the whole diffusion process with the predefined number of diffusion steps ($T$). For instance, we can define an arbitrary sampling trajectory with the subset of latent variables $\{x_{s_1}, \ldots, x_{s_T}\}$ that has a length smaller than $T$. Controlling the length of the sampling trajectory, it is able to search an acceptable number of sampling steps considering both the computational cost and the sample quality (which is discussed further in section 3.2 ).

2.2 Datasets and model architecture

2.2.1 Generation of synthetic microstructure

To train the diffusion probabilistic models for generating synthesized microstructures, a sufficient number of microstructure samples with different spatial variations must to be obtained. Furthermore, various types of microstructures should be considered to validate the effectiveness of diffusion models as a universal method for microstructure synthesis. Therefore, the following seven types of microstructures are considered in this study: 1) alloy, 2) carbonate, 3) ceramics, 4) copolymer, 5) polymethyl methacrylate, and 6) fiber composites. To aid the validation of the proposed approach, the microstructure images for training are generated using the descriptor-based microstructure reconstruction methods presented in the previous literature [3, 4, 52]. In addition, the microstructure samples are generated to have certain morphological features such as grain size distribution, two-point correlation functions, and volume fraction. Then, the models are tested to confirm if these features are well reproduced. For each type of microstructure, 1,000 images of size $64 \times 64$ are created. The detailed explanation for the reconstruction methods is provided in Appendix A. The sample images for the different kinds of reconstructed microstructures can be seen in Appendix B. In addition, 1,000 microstructure images of size $256 \times 256$ are also created for each of the three types of microstructure (alloy, carbonate, and fiber composites) to verify the generation performance on higher resolution.

2.2.2  Model architecture

Following the UNet[53]-based architecture presented by Ho et al. [41], which also resembles the unmasked PixelCNN++[54], is employed for microstructure reconstruction. The models have five feature map resolutions from $64 \times 64$ to $4 \times 4$ with the 128 base channels and have three convolutional residual blocks per resolution level. A global attention layer is used at the $16 \times 16$ level between the convolution blocks with a single head. The ablation study is not conducted yet which leaves room for further improvement of the diffusion models for MCR applications. It is also worth noting that the performance of the diffusion probabilistic models can be improved by increasing the number of attention heads while using multi-resolution attention as shown in the recent work proposed by Dhariwal and Nichol [38].

2.3 Evaluation metrics

To evaluate the quality of the generative samples with the diffusion models, the FID score is utilized since it has been the most widely used metric for image generation tasks. The FID assesses the quality of generated images by calculating the magnitude of difference between original and generated distributions using the pre-trained Inception v3 neural network model (often prepared on ImageNet datasets) [48, 55]. FID score can be computed as follows:

$$\|\boldsymbol{\mu}_r - \boldsymbol{\mu}_g\|_2^2 + \mathrm{Tr}\left(\mathbf{C}_r + \mathbf{C}_g - 2(\mathbf{C}_r \cdot \mathbf{C}_g)^{1/2}\right) \qquad (18)$$

where $\boldsymbol{\mu}$ and $\mathbf{C}$ are the mean vectors and the covariance matrices in the embedding space of Inception v3 without the classification layer (subscripts $r$ and $g$ real and generated data, respectively). In addition, FID has been known to be in good agreement with human perceptual scores [56, 57]. However, a significant drawback of FID is that it cannot separate the two critical components for evaluating the quality of the generative samples: fidelity and diversity. Thus, the improved precision and recall metric proposed is also employed to assess the performance of diffusion models. According to Kynkäänniemi [49], the precision and recall metric for generative models can be computed as follows:

$$Precision(\mathbf{\Phi}_r, \mathbf{\Phi}_g) = \frac{1}{|\mathbf{\Phi}_g|} \sum_{\phi_g \in \mathbf{\Phi}_g} f_{PR}(\phi_g, \mathbf{\Phi}_r) \qquad (19)$$

$$Recall(\mathbf{\Phi}_r, \mathbf{\Phi}_g) = \frac{1}{|\mathbf{\Phi}_r|} \sum_{\phi_r \in \mathbf{\Phi}_r} f_{PR}(\phi_r, \mathbf{\Phi}_g) \qquad (20)$$

where $\phi$ denotes the feature vector of the data (real/generated), and $\mathbf{\Phi}$ is the set of the feature vectors (real/generated). $f_{PR}(\phi, \mathbf{\Phi})$ is a binary function to determine whether the data of interest $\phi$ is located in the volume created by the feature vectors of the sets $\mathbf{\Phi}$ [49].

To verify the generated microstructure images in terms of morphological features, the conventional two-point correlation function [8] and the lineal path function [58] are also used to analyze the morphological differences between the original and the generated microstructure images. The two-point correlation function can be defined as follows:

$$S_2^{(i)}(\mathbf{r}_1, \mathbf{r}_2) = X(\mathbf{r}_1)X(\mathbf{r}_2) \qquad (21)$$

where $\mathbf{r}_1$ and $\mathbf{r}_2$ are randomly selected location vectors and $X$ is a binary function that returns 1 if there is a phase of interest $i$ at a given location and 0 otherwise. It is worth noting that $S_2^{(i)}$ only depends on the Euclidean distance between $\mathbf{r}_1$ and $\mathbf{r}_2$ [59]. Similar to the two-point correlation function, the lineal path function can be computed to evaluate the connectivity between the clusters of phase $i$ as follows:

$$L^{(i)}(\mathbf{r}_1, \mathbf{r}_2) = \begin{cases} 1 & \text{if a linear line between } \mathbf{r}_1 \text{ and } \mathbf{r}_2 \text{ lands on phase } i \\ 0 & \text{otherwise} \end{cases} \quad (22)$$

Then, the discrepancy between the correlation functions (Table 3 and Table 4) is computed as follows to quantify the difference between the original and generated microstructure images

$$\varepsilon^c = \frac{A_c}{A_{ori}} \times 100 \quad (23)$$

where $A_c$ is the area between the two correlation functions, and $A_{ori}$ is the area under the correlation function of the original microstructure images.

Furthermore, it is essential to check if the models are overfitting the training dataset. The most common form of overfitting in generative models is that the models output only the training samples, also called data-copying [60]. To verify the performance of generative models, the models should be tested if the data-copying has occurred (section 3.1.3). Thus, the $n \times m$ size generated microstructure images closest to the training images are assessed in terms of the mean squared error of pixel values computed as

$$\varepsilon_{pixel} = \frac{1}{nm} \sum_{i=1}^{n} \sum_{j=1}^{m} (Y_{ij} - \hat{Y}_{ij})^2 \quad (24)$$

where $Y_{ij}$ and $\hat{Y}_{ij}$ are pixel values of training and original images, respectively.

2.4 Training procedure

In this study, the diffusion-based generative models for microstructure reconstruction are trained using Adam optimizer [61] with a learning rate of 0.0001, training steps of 100000, and a batch size of 32 per GPU. The exponential moving average (EMA) rate of 0.9999 is used with a dropout ratio of 0.1 for all experiments. The diffusion models are implemented with PyTorch library [62] and trained on Nvidia RTX A6000 GPUs. The diffusion time steps for all models are set to be 4000 with the linear noise schedule.

## 3. Results and discussion

3.1 Model performance for microstructure reconstruction

3.1.1  FID scores

The randomly selected generated microstructure images using the diffusion models with 150 DDIM sampling steps are shown in Figure 2 (more samples can be seen in Appendix C). As shown in the figure, the diffusion models can generate visually similar microstructures compared to the original microstructure images. To quantify the performance of the generative models, the evaluation metrics (FID, precision, and recall) are computed for each type of generated microstructure sample as shown in Table 2. It can be found that the lowest FID score is achieved for the polycrystalline alloy microstructure images with the highest precision and recall (0.784 and 0.758, respectively). This could be explained by the relatively simple geometrical structure of the considered alloy images consisting of only the straight grain boundaries (i.e., white pixels). Since Song et al. [40] achieved the FID score of 4 on CelebA ($64 \times 64$) datasets with the DDIM approach, the observed FID scores in this study are not remote from the state-of-the-art generative models. In addition, Dhariwal et al. [38] also achieved the state-of-the-art FID score of 2 on ImageNet ($64 \times 64$) datasets with their proposed guided diffusion approach surpassing GANs. However, since the FID score is computed with the pre-trained Inception V3 model [48, 55], which means the score is calculated within the latent space of the network trained for image recognition, it is not easy to say that the proposed models are well-trained for the microstructure reconstruction tasks. In other words, the FID scores show that the images are visually alike regarding human perception of similarity. Thus, the detailed morphological features of the generated microstructure samples are also evaluated in terms of the spatial correlation functions [8, 58] in the next section.

3.1.2  Morphological features assessed with correlation functions

The comparison between the two-point correlation and the lineal path functions of the original and generated microstructure samples are as depicted in Figure 3. It is worth noting that the deviations of the functions, which are depicted with the shaded envelope in the graphs, are quite small since the original samples are generated with the descriptor-based MCR methods (Appendix A) to have certain morphological features. To verify that the diffusion models can reproduce these features, the values of the discrepancy between the original and generated samples are computed as shown in Table 3 and Table 4. For all the cases, the discrepancy values are lower or about 5%, which shows that the generated samples are in good agreement with the original samples. Furthermore, the results show that the volume fractions are well preserved in the generated samples since the values of the correlation functions (original/generated) are almost the same when the distance between the randomly selected points (r) is 0. In addition, relatively high discrepancy rate values for the alloy, carbonate, and polymethyl methacrylate are observed as shown in Table 3. This can be explained by the relatively low values of the correlation functions for these materials, as shown in Figure 3 (which also means low volume fractions) where a slight difference leads to a significant increase in the discrepancy rate. In other words, the original and generated samples have visually similar morphologies (section 3.1.1) and near statistically equivalent microstructures in terms of spatial correlations.

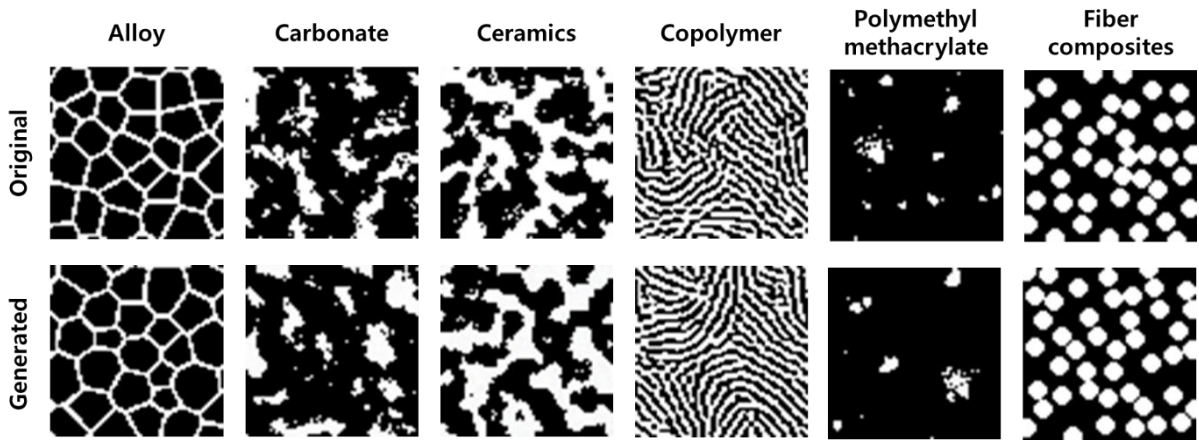

**Figure 2.** Comparison of the original microstructure images and the synthesized microstructure images using the trained diffusion model with DDIM-based sampling

**Table 2. FID scores evaluated on different types of microstructures (resolution of 64 × 64) with trained diffusion models**

| Microstructure | FID Score | Precision | Recall |
|---|---|---|---|
| Alloy | 2.08 | 0.784 | 0.758 |
| Carbonate | 18.12 | 0.324 | 0.712 |
| Ceramics | 13.87 | 0.384 | 0.625 |
| Copolymer | 6.74 | 0.483 | 0.784 |
| Polymethyl methacrylate | 11.29 | 0.637 | 0.791 |
| Fiber composites | 2.22 | 0.76 | 0.782 |

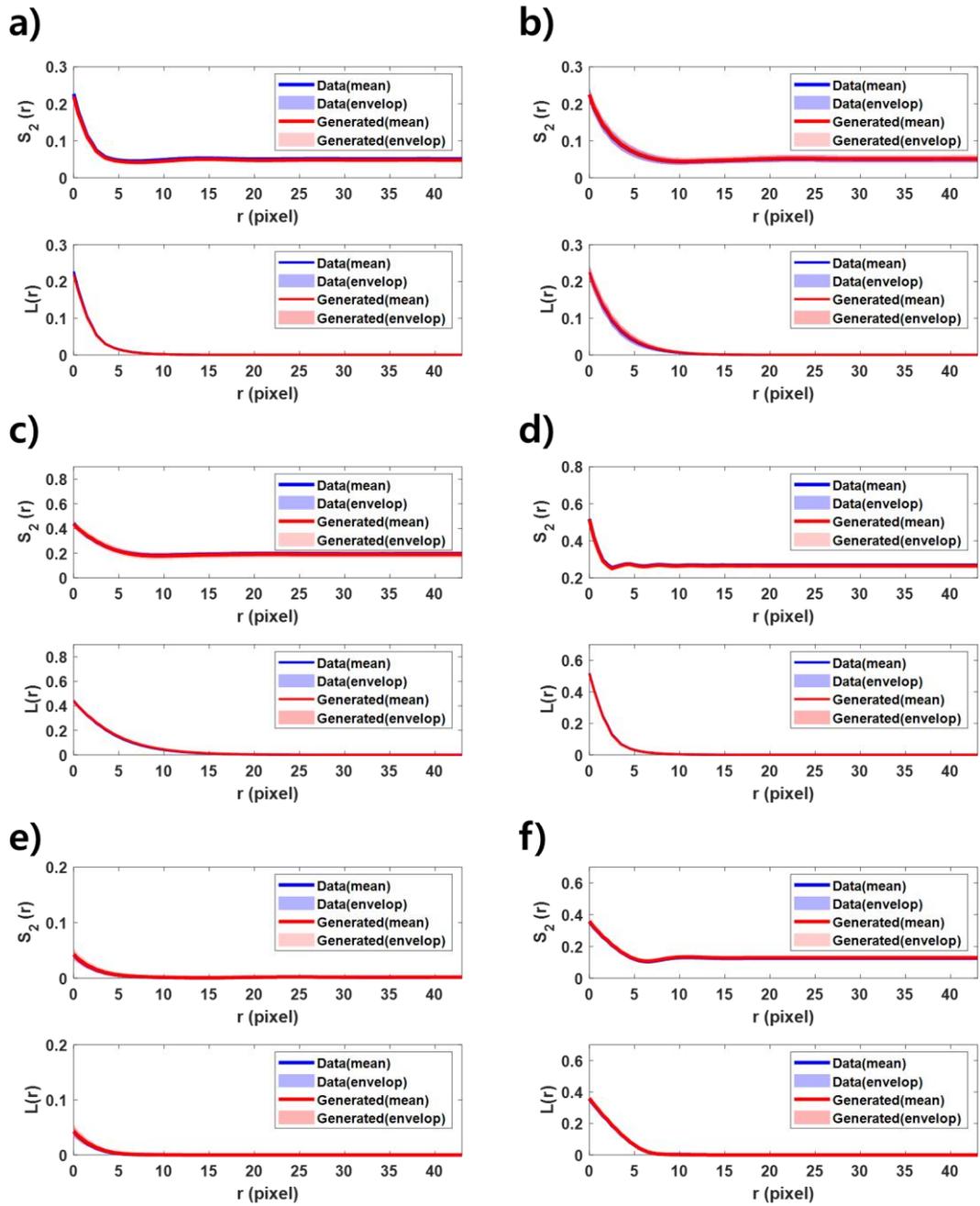

Figure 3. Comparison of two-point correlation functions between the original microstructures and the synthesized microstructures using the diffusion models: a) alloy, b) carbonate, c) ceramics, d) copolymer, e) polymethyl methacrylate, and f) fiber composites

**Table 3. The discrepancy between the two-point correlation functions ($S_2$) for different microstructures**

| Microstructure | The discrepancy between the correlation functions $\varepsilon_{TP}^c$ (%) |
| --- | --- |
| Alloy | 5.17 |
| Carbonate | 2.02 |
| Ceramics | 1.05 |
| Copolymer | 0.66 |
| Polymethyl methacrylate | 3.70 |
| Fiber composites | 1.30 |

**Table 4. The discrepancy between the lineal path correlation functions ($L$) for different microstructures**

| Microstructure | The discrepancy between the correlation functions $\varepsilon_L^c$ (%) |
| --- | --- |
| Alloy | 0.32 |
| Carbonate | 5.36 |
| Ceramics | 1.82 |
| Copolymer | 1.08 |
| Polymethyl methacrylate | 4.71 |
| Fiber composites | 0.81 |

Moreover, it is crucial to check whether the generated microstructure samples have similar physical characteristics compared to the original ones. In particular, the grain size is directly related to the tensile strength of the polycrystalline alloys (i.e., Hall-Petch relationship [63]). To validate the generated samples of alloy microstructure in terms of the physical properties, the grain size distributions are evaluated as shown in Figure 4. As shown in the figure, the distributions are very similar as well as the means ($\mu_{\text{mean}}$) and the standard deviations ($\sigma_{\text{dev}}$). Regarding this result, it can be said that the diffusion models are able to reconstruct visually similar alloy microstructure from the given datasets while capturing the grain characteristics which are closely related to the material properties. However, it is still required to validate the practical application of the diffusion models with various kinds of alloy features such as high grain aspect ratios (i.e., columnar structures) and abnormal grain growth [64]. Furthermore, training diffusion models with the inverse pole figure (IPF) maps considering the relationship between the adjacent grains (e.g., disorientation angle) could be conducted in the future.

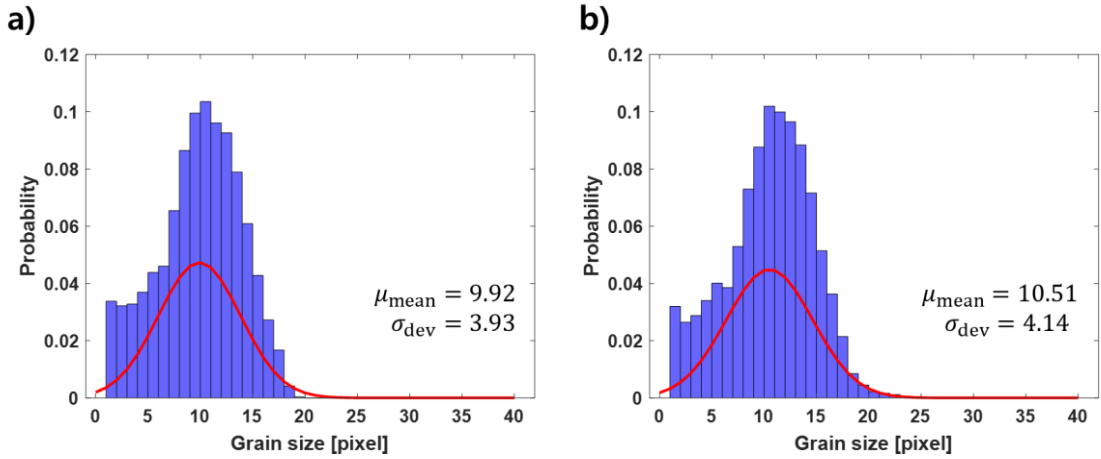

**Figure 4. Comparison of the grain size distribution between a) the original alloy microstructure samples and b) the generated alloy microstructure samples**

In particular, the results also show that the trained diffusion models can generate visually similar copolymer microstructures (Figure 2). According to the previous literature [5], the Gaussian random field[10, 65], decision tree [12], physical descriptor [7], and two-point correlation [18] based reconstruction methods cannot generate visually similar copolymer microstructures although they may achieve low discrepancy in terms of n-point correlation functions. This is due to the difficulties of characterizing the unique feature of copolymer microstructure, the local anisotropy along with global isotropy. On the contrary, the generated copolymer microstructure images with the diffusion models have various locally anisotropic structures (Figure 5 and Appendix C) while retaining the global isotropy. Furthermore, the connectivity of the white phase is reproduced well in all the generated samples. The FID score of 6.74 with low discrepancy rates of correlation functions (0.66% and 1.08% for $S_2$ and $L$, respectively) also show that the generated copolymer samples are visually and statistically similar to the original samples. Notably, this highlights the performance of the diffusion models as a potential universal microstructure reconstruction tool since the method does not require any additional modifications for generating complex microstructure images.

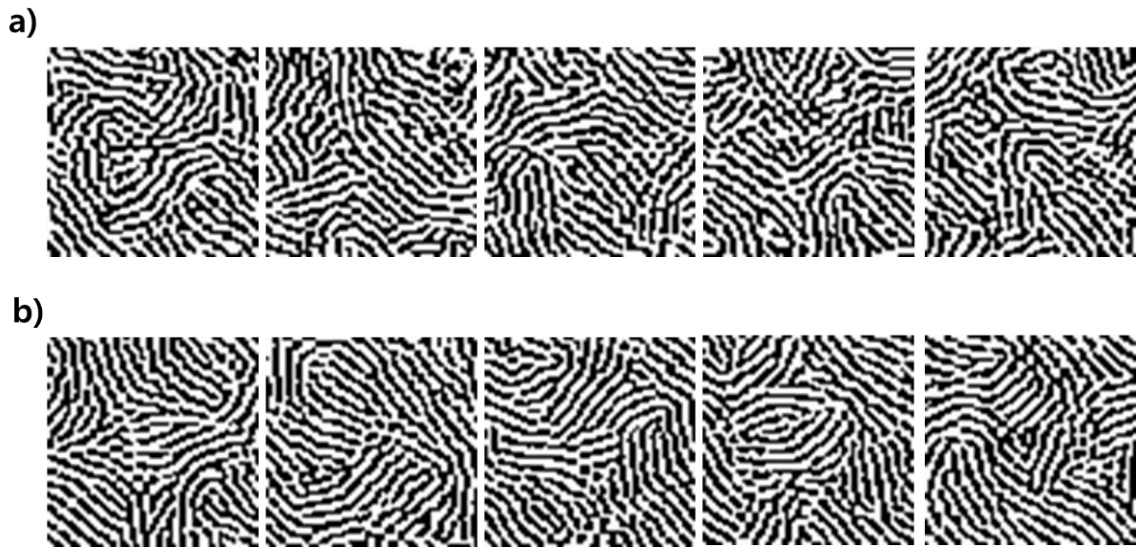

**Figure 5. Comparison between a) original copolymer images and b) generated copolymer images with the diffusion model**

It is also worth noting that the circular structures of the fiber composites are successfully reproduced with the diffusion models (Figure 2 and Appendix C). Since fixed morphologies of rectangular and circular structures are often distorted in the samples reproduced with generative models such as GANs [66-69], additional regularizations are often needed to capture these features. However, the results in this study are obtained without any additional regularizations for the morphology control. Although more various kinds of fixed morphologies such as rectangles, ellipsoids, and lattices, need to be considered in the future, the results imply that the diffusion model is a possible solution for the reconstruction of the fixed morphological features.

In addition, a relatively high FID score of 18.12 is obtained for the case of generated carbonate microstructure samples as shown in Table 2. This is probably due to the samples' complicated structures including some small dispersed materials (i.e., white pixels). The lower precision value of 0.324 compared to those of the other types of microstructure also means that the fraction of the realistic generated images is small. Meanwhile, the discrepancy rates of 2.02 and 5.36% (Table 3) for the correlation functions show that the generated samples have near statistically equivalent microstructures compared to the original samples. According to Li et al. [5], the discrepancy for the case of carbonate microstructure was found to be about 3~7% using the transfer learning method, decision tree, Gaussian random field, two-point correlation, and physical descriptor-based method. Regarding this result, it seems that the observed discrepancy rates indicate an acceptable performance for reconstructing carbonate microstructure.

3.1.3   Evaluation of pixel error for checking data-copying

To detect data-copying in the generated samples, the generated samples that have minimum $\varepsilon_{\text{pixel}}$ are illustrated with the corresponding training images as shown in Figure 6. As $\varepsilon_{\text{pixel}}$ gets close to zero, the images are near identical in the pixel space. It can also be seen that the minimum $\varepsilon_{\text{pixel}}$ increases with the fraction of the white pixels as there is more chance of discrepancy in pixel values. Since all the generated images that have minimum $\varepsilon_{\text{pixel}}$ values are different from the corresponding training images, it can be said that data-copying has not occurred during the generation process.

In particular, the generated alloy image which has the minimum $\varepsilon_{\text{pixel}}$ shows an interesting characteristic of the diffusion models. As shown in Figure 7, the generated sample has two distinctive regions. The region enclosed by the blue box has high $\varepsilon_{\text{pixel}}$ whereas the region enclosed by the red box has low $\varepsilon_{\text{pixel}}$ with respect to the original training sample. In other words, the region in the red box can be thought of as a copy of the original sample shown in Figure 7a. However, the generated sample is obviously different from the original sample since the blue-boxed region has different grain structures (high $\varepsilon_{\text{pixel}}$). It is worth noting that the regions in the blue and red boxed are well-connected, forming the grain boundaries (i.e., white pixels). Although the models are trained without any morphological control or additional regularization, the generated image shows that the connection of grain boundaries is preserved.

That is, naively speaking, the diffusion models are seemed to learn the essential features of grain structures. It emphasizes the performance of the diffusion models as a potential universal MCR method for preserving important microstructural features. However, it is required to study the trained models' semantic latent space in future work to ensure whether the diffusion models can fully capture various types of complex microstructural features. Meanwhile, the generation results show that the trained diffusion models can reconstruct visually similar and statistically equivalent microstructure images that are not present in the training dataset.

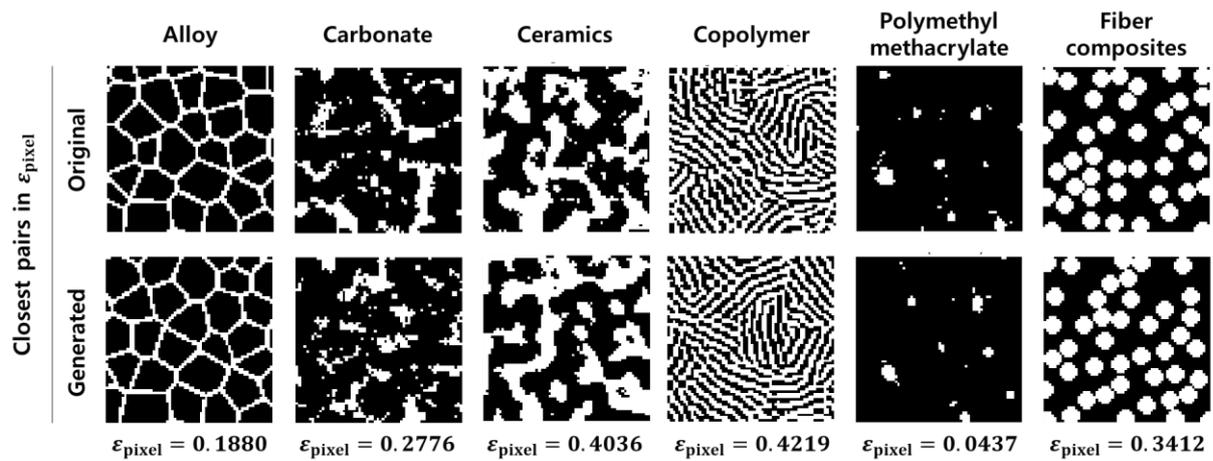

**Figure 6. The original microstructure images and the corresponding closest synthesized microstructure images in terms of the mean squared error of pixel values**

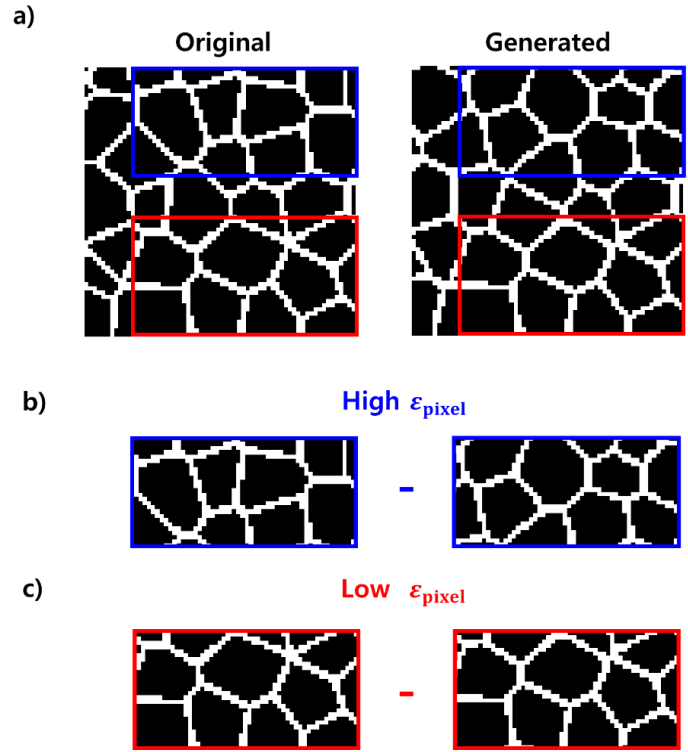

**Figure 7. Comparison of original and generated samples (alloy) that are closest in terms of the mean squared error of pixel values: a) corresponding samples, b) region of high pixel error, and c) region of low pixel error**

3.2 Effect of sampling steps

It has been known that the quality of the generated samples with DDIM is highly affected by the number of sampling steps. For instance, Song et al. [40] showed that DDIM with 20 to 100 sampling steps could generate samples with quality comparable to DDPM with 1000 diffusion steps. To analyze the effect of the sampling steps, the evaluation metrics are computed with the different number of sampling steps for $64 \times 64$ size carbonate microstructure samples as shown in Figure 8a. The graph shows that the FID score is very high, above 40 with 10 sampling steps, and significantly decreases as the number of steps increases to 150. With 300 sampling steps, the FID score seems to converge to a specific value of about 17.46. The precision and recall values also reach their maximums with 300 sampling steps. Meanwhile, the sampling time for the generation of 10,000 images almost linearly increases with the length of the sampling trajectory (Figure 8b) as explained in section 2.1. Since there is a trade-off between sampling time and sample quality, it is essential to determine optimal DDIM sampling steps to acquire enough samples considering computational capacity.

To verify the performance of the diffusion models on the reconstruction of higher resolution microstructure images, the diffusion models are trained with the $256 \times 256$ images of alloy, carbonate, and fiber composites (Appendix D). The generated samples with the trained diffusion models using the same training procedure (section 2.4) are shown in Figure 9 with 500 DDIM sampling steps (more samples are provided in Appendix E). As can be seen in the generated samples, the microstructural features, such as the connected grain boundaries and the circular fibers, are well reproduced. The evaluation metrics are also obtained with the different sampling steps, as shown in Figure 10a for the carbonate microstructure ($256 \times 256$). Similar to the case with $64 \times 64$ size images, the FID score increases with the number of DDIM sampling steps. In particular, the precision of the generated samples also keeps increasing with sampling steps up to 500. This implies that more sampling steps might be required to generate higher resolution microstructure images with acceptable quality. In summary, the number of sampling steps in the diffusion models needs to be determined considering sample resolution, sample quality (assessed with the evaluation metrics such as FID, precision, and recall), and computational capacity. Furthermore, the parameters including the number of diffusion steps and the hyperparameters of the network architecture may need further modifications to optimize the reconstruction performance for specific sample resolutions.

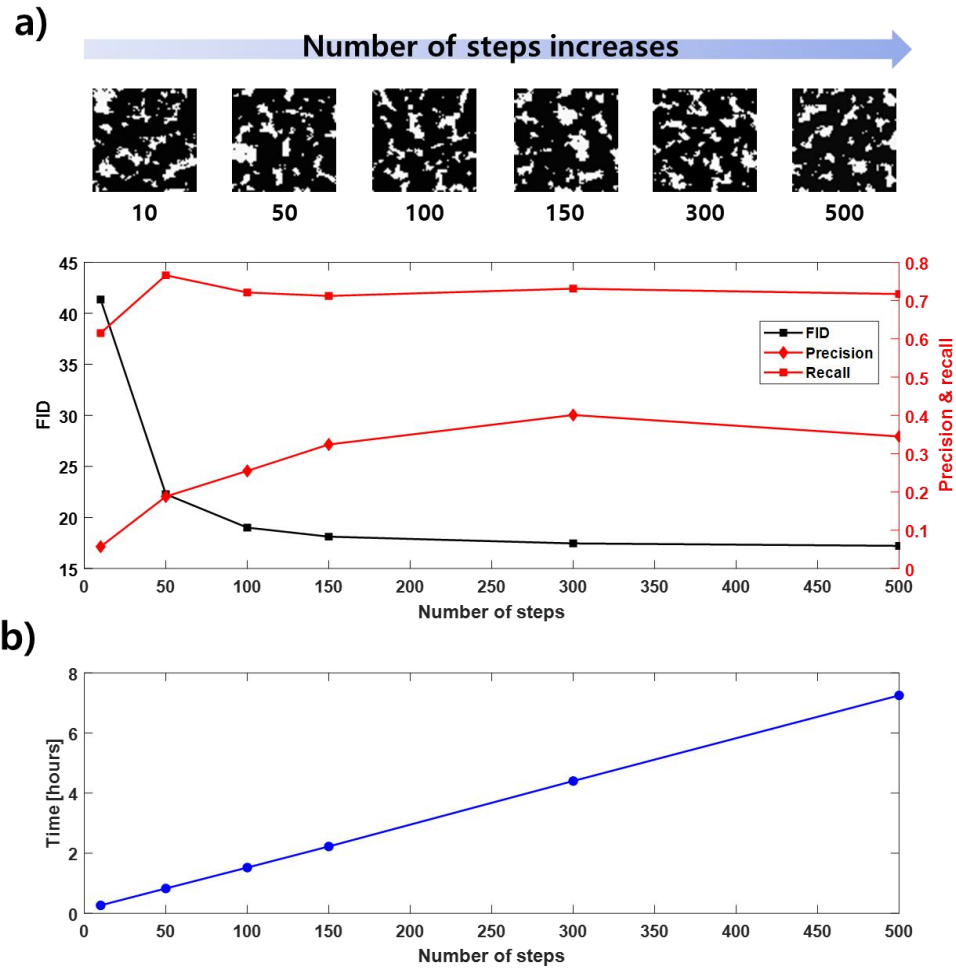

**Figure 8.** Effect of number of sampling steps for generation of $64 \times 64$ microstructure images: a) metrics (FID, precision, and recall) for evaluation of microstructure (carbonate) reconstructions, c) computational time to sample 1k images

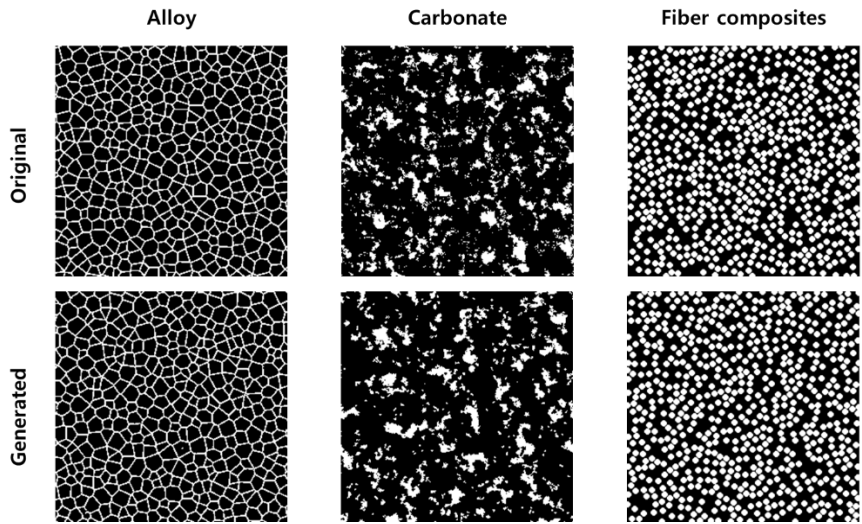

**Figure 9.** Comparison of the original microstructure images and the synthesized microstructure images using the trained diffusion model with DDIM-based sampling with 500 sampling steps

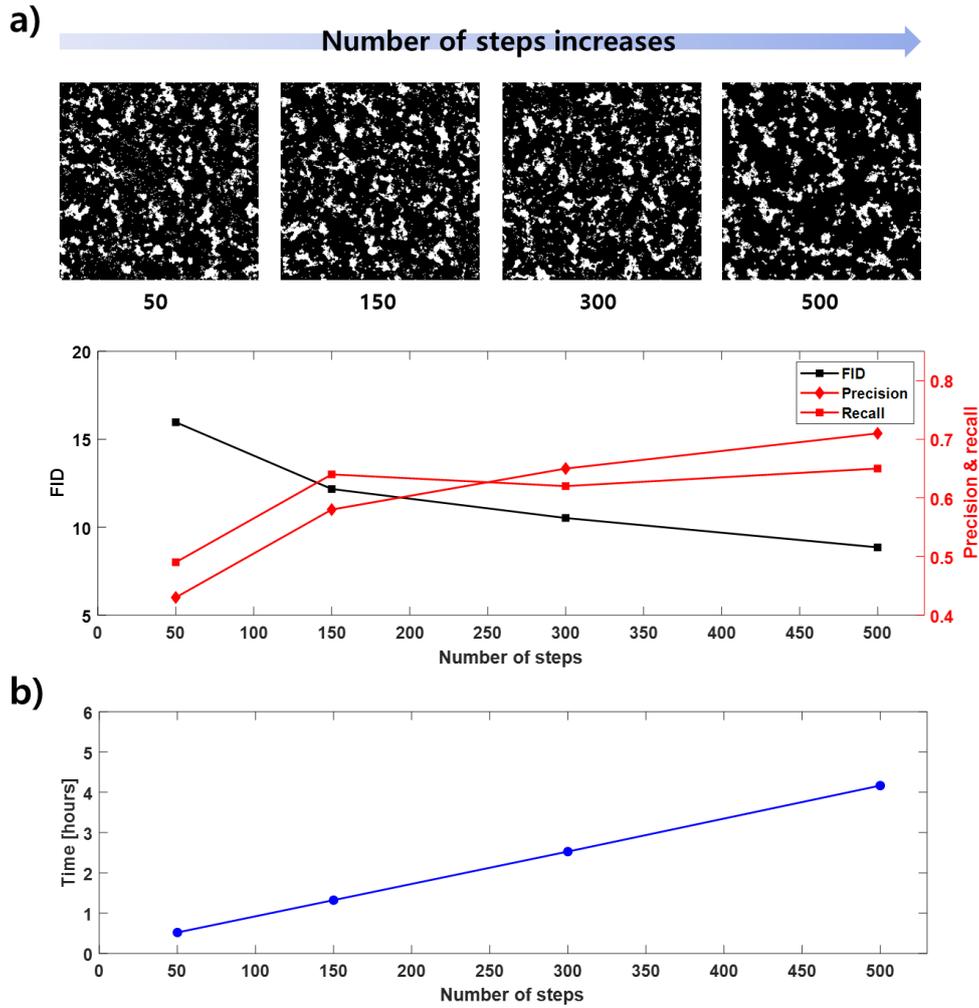

**Figure 10. Effect of number of sampling steps for generation of 256×256 microstructure images: a) metrics (FID, precision, and recall) for evaluation of microstructure (carbonate) reconstructions, c) computational time to sample 1k images**

## 4. Conclusion

In this study, the denoising diffusion models are first employed for the microstructure reconstruction tasks considering various types of microstructures. To verify the reconstruction performance of the models, the synthetic microstructure samples with certain morphological features (e.g., grain size, volume fraction, two-point correlation functions, and local anisotropic structures) are used as training datasets. The trained denoising diffusion models are then used to generate samples for each type of microstructure using DDIM sampling to accelerate the sampling process. The results show that the diffusion models can reproduce visually similar microstructure images and the quality of the samples is assessed with several metrics (FID score, precision, and recall). The generated samples also have similar spatial correlation functions with those of the original samples, indicating that the statistical distributions of morphological features are well reproduced. We suggest that the diffusion models must be further studied to develop a universal microstructure reconstruction tool due to its simple implementation, stable training, and the small number of parameters that need to be controlled.

The following potential types of future research for microstructure reconstruction with denoising diffusion models are recommended. To improve the reconstruction performance of the diffusion models, an ablation study can be conducted to find the best model configuration. Specifically, the parameters of the diffusion models (e.g., numbers of sampling and diffusion steps) and the hyperparameters (e.g., number of feature map resolutions, self-attention, and residual blocks) can be tuned for the specific microstructure reconstruction tasks. Furthermore, the semantic latent space of the diffusion models, which are the noised samples $\{x_1, ..., x_T\}$, need to be studied with the deterministic formulation of DDPM (i.e., DDIM). In order to control the microstructural features of the generated samples, manipulation techniques such as guidance and interpolation within the latent space need to be developed, which are getting much attention recently [70, 71]. The sampling speed needs to be also improved to aid the data-driven materials design with a sufficient amount of microstructure dataset.

**Acknowledgment**

This material is based upon work supported by the Air Force Office of Scientific Research under award number FA2386-22-1-4001. Any options, finding, and conclusions, or recommendations expressed in this material are those of the authors and do not necessarily reflect the views of the United States Air Force. The authors are grateful for their support

# Appendix

## A. Descriptor-based microstructure reconstruction

The considered types of microstructure are alloy, carbonate, ceramics, copolymer, polymethyl methacrylate, and fiber composites in this study. Firstly, the synthetic microstructure images of alloy are generated using MicroStructPy [52] with the given descriptors such as grain sizes and volume fractions $V_{f,i}$ to determine the population density. The probability $p_i$ that a certain grain is belonged to the corresponding phase $i$ can be calculated as

$$n_i = \frac{V_{f,i}}{\mathbb{E}[V_i]} \tag{25}$$

$$p_i = \frac{n_i}{\sum_j n_j} \tag{26}$$

where $n_i$ is the number of grains (phase $i$) per unit volume. The list of grains is generated with the probability distribution weighted by $p_i$ to determine the phase. Then, the positions of seeds are generated while matching the grain volume and checking if there is any overlap exceeds given tolerance. According to the assigned radius for each grain, a collection of circles can be obtained from the seed geometries. Finally, synthetic grain structures can be obtained after polygonal meshing using Voronoi tessellation and the collection of circles. In addition, three types of grains are considered with the effective size ratio of 1:2:2.5:3:4 with the volume fraction ratio of 0.15:0.2:0.3:0.2:0.15 in this study.

The synthetic microstructure samples of fiber composites are generated using the random sequential absorption (RSA) algorithm [72, 73]. In RSA algorithm, the circular particles are randomly positioned and added to the pre-defined domain sequentially. If the added particle intersects with the pre-existing particles, the location of the added particle is sampled again until there is no intersection. Then, this process is repeated until the target fiber volume fraction (which is set to be 0.35 in this study) is reached.

To generate synthetic microstructure images for the rest of microstructures, the descriptor-based microstructure reconstruction software platform of MCRpy [3, 4] is used. MCRpy provides features of MCR based on various types of descriptors including the volume fractions, the spatial correlations, Gram matrices [74], and normalized total variation. In this study, the spatial two-point correlation $S_2$ is used as a descriptor for generating training samples carbonate, ceramics, and polymethyl methacrylate. It was shown by Paul et al. [3] that the microstructures of these materials can be reconstructed well using only n-point correlation functions. Since it is able to formulate the objective function (i.e., root mean square error between $S_2$s of current and target microstructure images) with the differentiable $S_2$, the desired microstructure is obtained using truncated Newton (TNC) optimizer [75]. On the other hand, the microstructure of copolymer is hard to be reproduced with the spatial descriptor [3, 4]. Thus, the Gram matrices of the feature maps in the pre-trained VGG-19 convolutional neural network [74] is used to reconstruct the microstructure of copolymer in this study. In this case, the objective function is the mean square error between the Gram matrices of current and target microstructure images which is optimized with L-BFGS-B optimizer [76].

**B. Microstructure samples for training (64 × 64)**

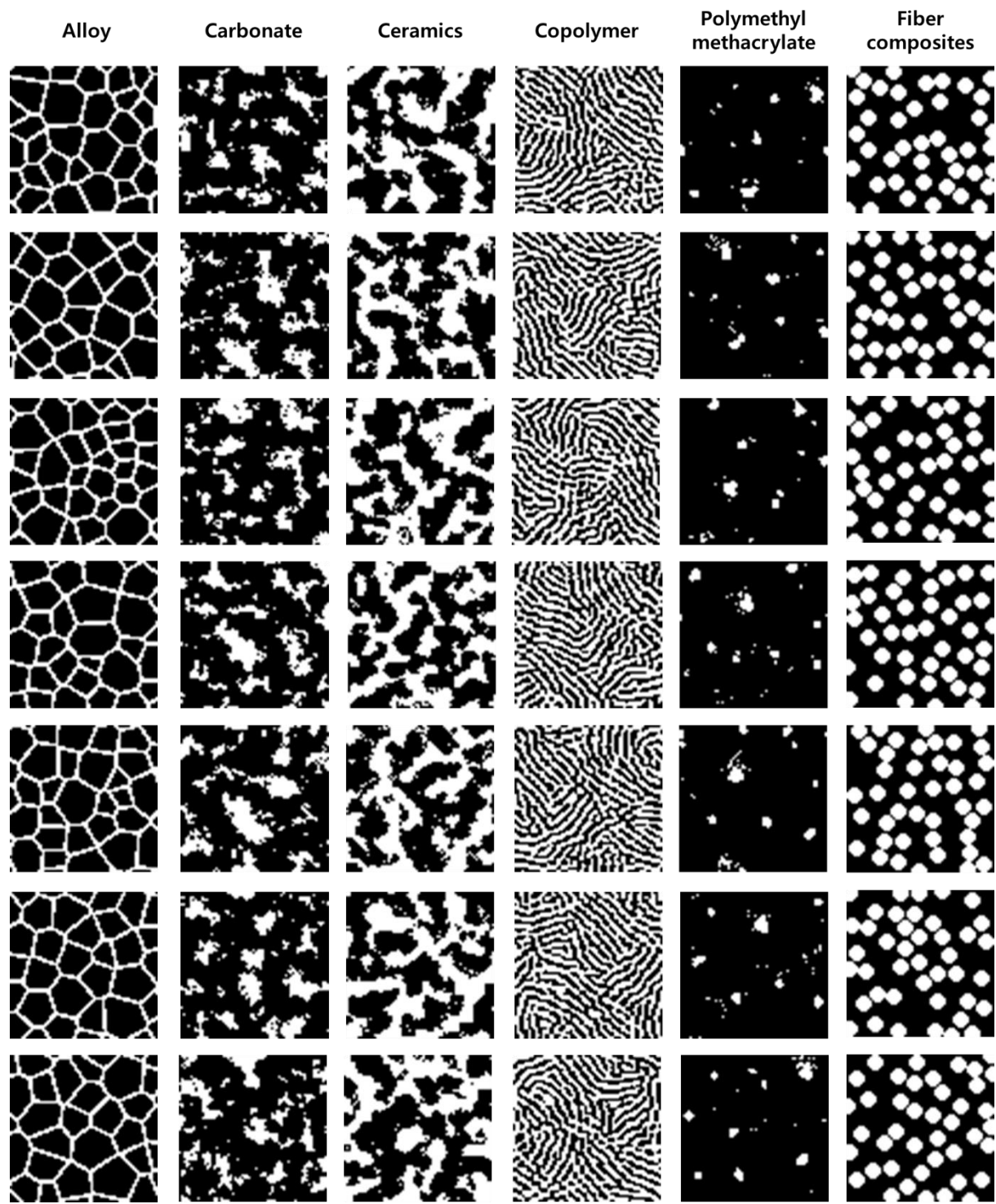

**Figure 11. Microstructure samples generated by descriptor-based microstructure reconstruction method for training the diffusion models (64 × 64)**

**C. Microstructure samples generated by diffusion model (64 × 64)**

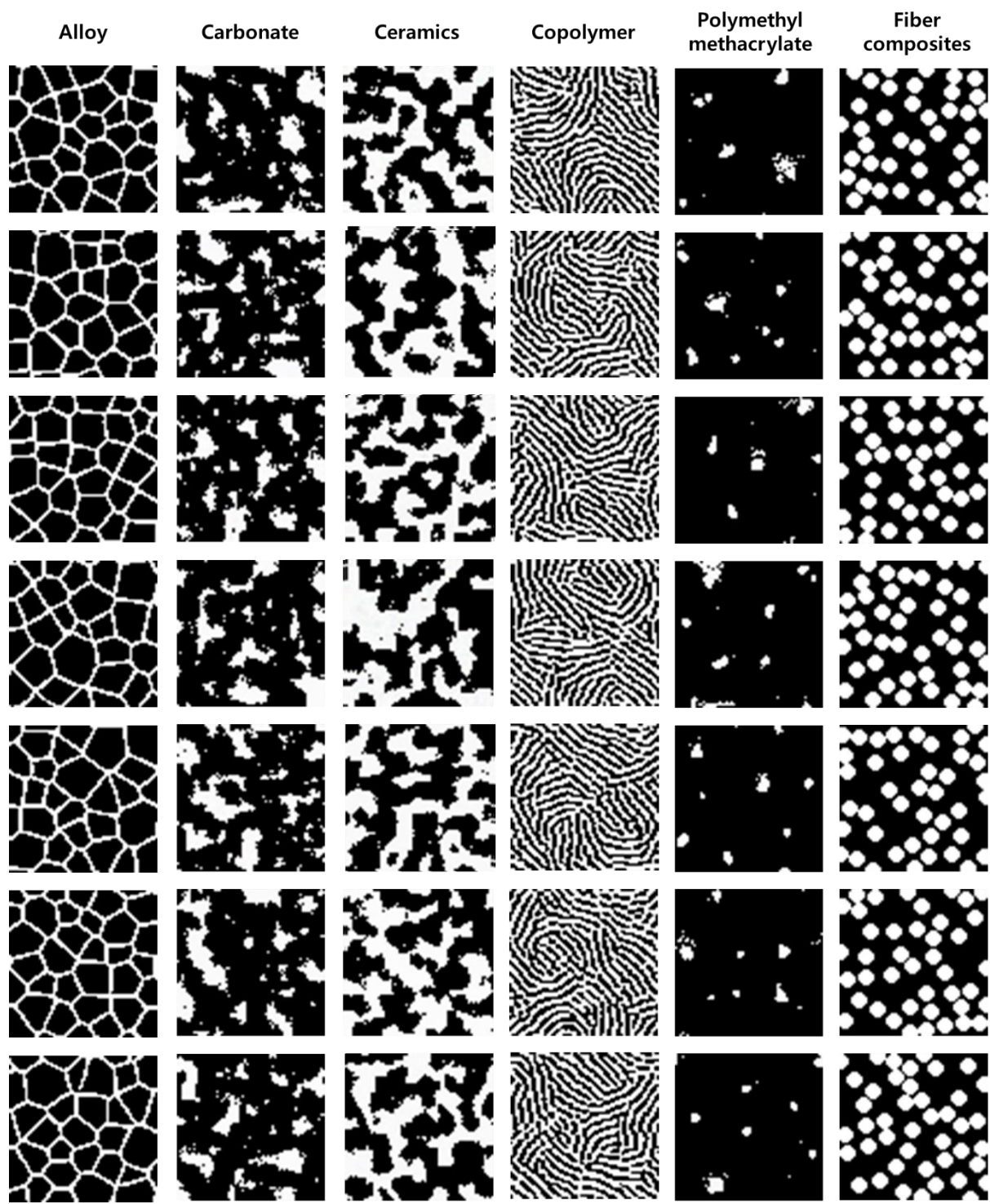

**Figure 12. Microstructure samples generated by the trained diffusion models with 150 DDIM sampling steps (64 × 64)**

## D. Microstructure samples for training (256 × 256)

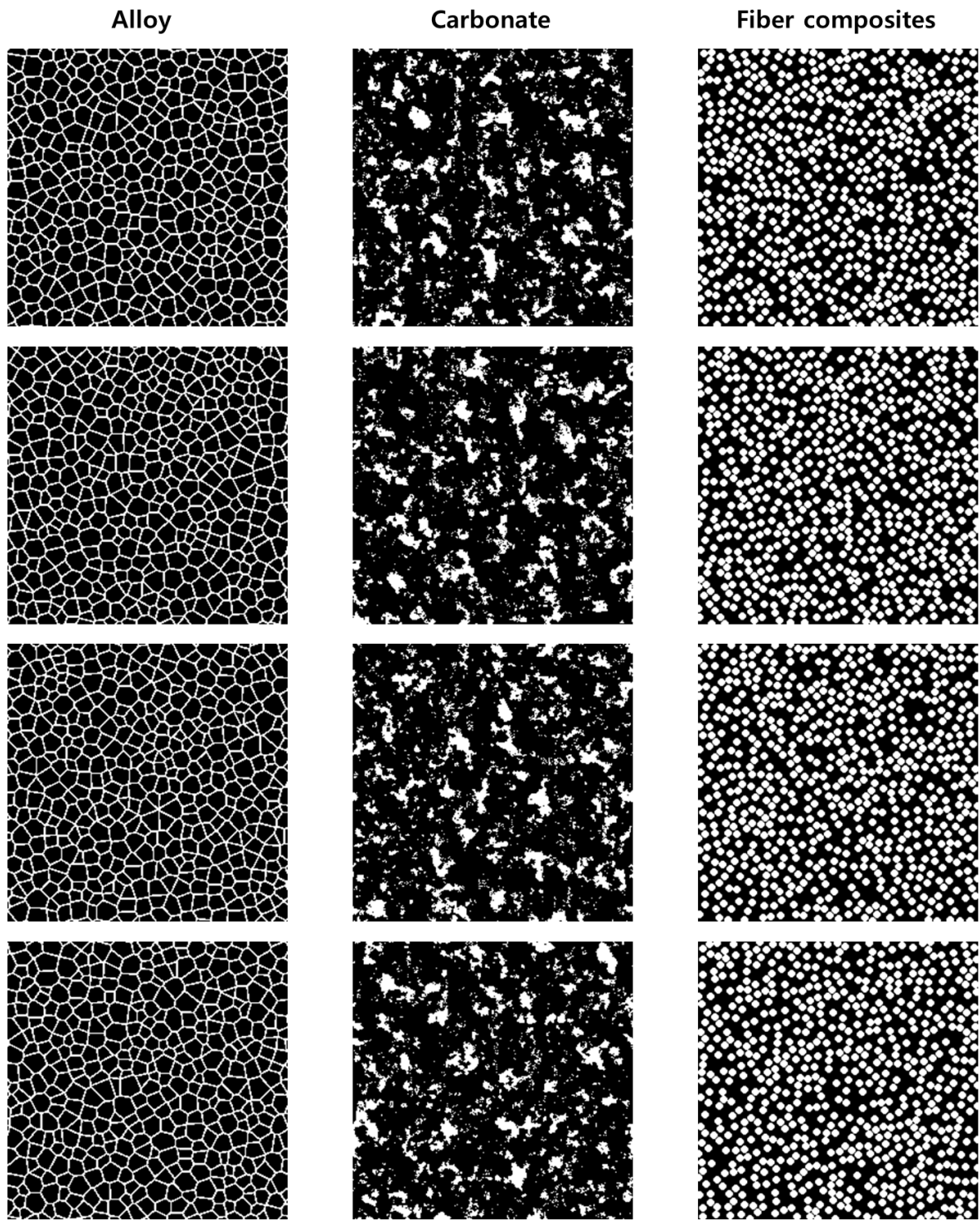

Figure 13. Microstructure samples generated by descriptor-based microstructure reconstruction method for training the diffusion models (256 × 256)

**E. Microstructure samples generated by diffusion model (256 × 256)**

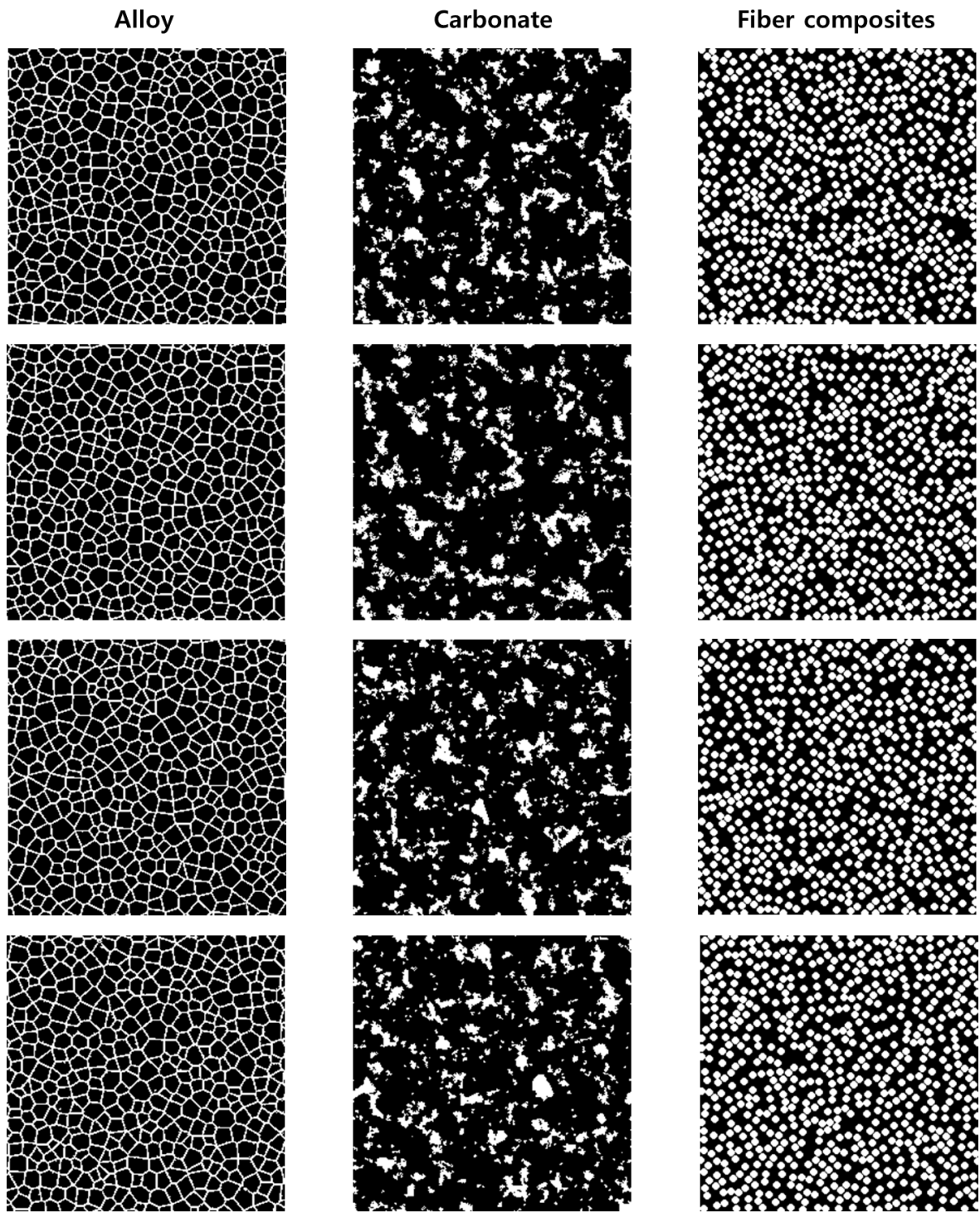

**Figure 14. Microstructure samples generated by the trained diffusion models with 500 DDIM sampling steps (256 × 256)**